# Personal Productivity and Well-being

By Jenna Butler, Mary Czerwinski, Shamsi Iqbal, Sonia Jaffe, Kate Nowak, Emily Peloquin, Longqi Yang

Chapter 2 of *The New Future of Work: Research from Microsoft into the Pandemic's Impact on Work Practices*, edited by Jaime Teevan, Brent Hecht, and Sonia Jaffe, 1st ed. Microsoft, 2021.

## TABLE OF CONTENTS

# 2 PERSONAL PRODUCTIVITY AND WELL-BEING

*By Jenna Butler, Mary Czerwinski, Shamsi Iqbal, Sonia Jaffe, Kate Nowak, Emily Peloquin, and Longqi Yang*

## 2.1 Introduction

We now turn to understanding the impact that COVID-19 had on the personal productivity and well-being of information workers as their work practices were impacted by remote work. This chapter overviews people's productivity, satisfaction, and work patterns, and shows that the challenges and benefits of remote work are closely linked. Looking forward, the infrastructure surrounding work will need to evolve to help people adapt to the challenges of remote and hybrid work.

## 2.2 Productivity, Satisfaction, and Work Patterns

*2.2.1 On average, self-reported productivity was unchanged, but it varied with role and experience.*

Since information worker productivity is hard to measure, many studies looked at self-reported productivity and satisfaction. Individually, some people flourished – and hoped to continue working from home at least part of the time post-COVID – but others struggled. In a March survey of US-based Microsoft software engineers and program managers, 34% said their productivity had decreased while working from home, and 34% said it had increased. In a July survey of employees in Puget Sound, 32% agreed that returning to the office would make them more productive and 30% disagreed [109]. Externally, we saw an initial information overload as people adjusted to new tools and ways of working [77]. Some people reported liking the new interpersonal communication patterns while others were hoping for a return to the "old way" [77].

Microsoft employees were more likely to report a decrease in productivity while working from home if they reported little prior experience with remote work, a shorter tenure at the company, fewer pre-pandemic collaborations (including meetings), or fewer standups. Also, software engineers were more likely than program managers to report their productivity being negatively affected [46]. Geographically, employees in Puget Sound were also more likely to report a decrease in productivity and lower satisfaction [46,128], potentially due to their having less experience with remote collaboration. In a survey of Microsoft employees in China, perceived personal and team productivity while working from home were significantly correlated with how one felt about physical health and mental well-being, perceived workplaces' setup and boundaries, meetings, team culture and workplace status.

Externally, in a spring 2020 survey of US workers commissioned by Stanford, 35% of respondents said they could be fully effective at their jobs while working remotely, 28% said they could do their jobs with 50 to 90% efficiency, and 37% said that they could not do their jobs at home or would be less than 50% efficient [18]. A pre-pandemic study [66] that used Occupational Information Network (O*NET) data on job content estimated that 37% of jobs (representing 46% of wages) could be done remotely [38]. In addition to job characteristics, individual circumstances mattered: a study of information workers found that whether they felt they could work from home productively depended on 1) the layout of their workspace and 2) whether they lived with others. Those living with others found their productivity hinged upon the needs of their co-habitants – children, partners needing help with errands or tasks, etc. [156].

These results are in line with prior remote work experiments. Nick Bloom and colleagues ran an experiment with call center workers who had expressed interest in working from home and found that, while the average effect of doing so was positive, some people were negatively affected. Those who were less productive at home were more likely to return to the office after the experiment ended [17]. A different experiment at the US Patent Office also found positive average effects, particularly for those who chose to move to areas with lower cost of living [32].

Prior research also found that factors like personality [125] and gender [1,63] played roles in remote work outcomes. Previous work also suggests that work that is easily codified is more amenable to remote work [14,168], but that creative work [59], new workstreams [53,140], and tasks that require extensive collaboration will suffer when done remotely [59,168]. Connecting these findings to the roles and types of work done at Microsoft is an important topic for future research.

While it is hard to measure productivity objectively, there are available objective measures for short-term developer activity; see the following chapter on [Software Engineering Experiences](#) for more details. On average, pull requests stayed constant or increased slightly. However, according to an internal study [69], the number of pull requests per new hire that started following the post-COVID move to remote work were 34% lower compared to pull requests per new hire during the same period in 2019; a greater percentage of developer new hires had not completed any pull requests in the first 90 days and the median time to complete a first pull request increased by 28% relative to 2019 [69]. New hires may not know the team as well, they may lack context, and it might be especially hard for managers to manage them remotely. The same study found that vendors completed 27% fewer pull requests on average this year compared to 2019.

*2.2.2 While some reported satisfaction with remote work, it varied with organization, role, and experience.*

Just as productivity varied extensively by role and experience, so did job satisfaction. In a survey of US-based Microsoft software engineers and program managers, 31% said that their job satisfaction had decreased, but 32% said it had increased [46]. Similarly, 33% of Microsoft Puget sound employees responding to a survey agreed that returning to the office would improve their overall well-being while 29% disagreed [109]. In a CSEO (Core Services Engineering and Operations) survey in November 2020, 81% of global Microsoft employees were satisfied (12% dissatisfied) with work from home [128]; a study in OXO (Office Experience Organization) yielded similar results, with over 60% reporting they're satisfied working from home [26].

Employees were more likely to be satisfied with working from home during COVID-19 if they had experience working from home prior to the pandemic [46,128]; in the CSEO survey 85% of global Microsoft employees who worked from home weekly before COVID-19 were satisfied working from home during COVID-19 (compared to 69% of those who never worked from home previously). This is also consistent with the results of an OXO survey, in which 61% of respondents said the challenges they faced early on improved over time [26]. Some people who worked remotely prior to COVID-19 reported that the totally-remote meetings they had in spring 2020 worked better than meetings pre-COVID when they were remote and everyone else was in a conference room [163].

People's satisfaction was correlated with their feelings of commitment, motivation, focus and being overworked. Nearly one-third (31%) of respondents to an external survey who felt committed to their team goals preferred working from home over the office, compared with 18% of those who felt disconnected from their team's objectives [180]. In the OXO study [26], people were less likely to report being satisfied with their work on days they mentioned challenges with motivation, focus, or feeling overworked. The same study found that mental and physical health, along with motivation and feeling overworked, were strongly associated with someone reporting being satisfied on less than 60% of their nightly responses. These challenges and the relationship to job satisfaction were not specific to working from home but could have been exacerbated by it. Managers in OXO were less likely than individual contributors to at some point report challenges related to motivation and collaboration, but more likely to have challenges around meetings, kids, and well-being [26].

A study from eXperience Collective Planning & Research [156] proposed five common remote work "journeys" based on feedback collected from external information workers, a framework that could help categorize the aforementioned experiences. The first category was the *Positive* experience, where people felt well-prepared for the challenge and embraced the benefits that it presented. The second category, *Growing*, represented the experiences of those who struggled initially, but then adapted and ended up feeling good. The third group, *Resilient*, represented those whose productivity and attitude towards remote work varied day by day. The fourth group, *Struggling*, included people who initially enjoyed the benefits of remote work, but later struggled as negative effects of isolation and distraction crept in. Finally, the last journey, *Negative*, represented those who felt overwhelmed with the challenges from the beginning and continued to have a negative experience. Of note here, the individual's particular "journey" was very much impacted by their job demands, workplace setup and whether they cohabitated with other people.

*2.2.3 Average workdays got longer during COVID.*

On average, the workday during COVID started earlier and ran later [22,26,66,128,151]. One internal study found that seven out of ten people experienced workweeks that expanded by at least three hours, that many people started to skip the lunch hour, and that the share of instant messages sent between 6 pm and midnight increased by 52% on average [3,66]. A study of external Teams users showed an 100% uptick in instant messages on Teams, particularly after hours [6]. GitHub data on developer activity also showed a lengthening of the workday for non-Microsoft employees [47]. In a study of information workers external to Microsoft, people reported that they found it more difficult than expected to stop working at the end of the day [156]. Internally, the change of workdays varied by roles [182]; sales and marketing professionals experienced shorter work days, and the changes for IT and researchers were not significant.

We see a similar pattern globally, potentially exacerbated by time zone differences. When Microsoft China offices switched to remote, average workday length (the time between the first and last work activity) increased. When they began to return to offices, it lengthened still more [131]. In a study of India-based Microsoft employees, it was found that the greater the time difference with collaborators, the bigger the challenge in managing the length of the workday [83]. Many people spent much of the day on other responsibilities – such as parenting or taking care of the household – and worked in the evenings.

Work also started extending into weekends more than it had before COVID-19. Using SwiftKey data as a proxy for mobile users, Microsoft Customer Insights Research (MMX) analyzed how usage of Teams and Outlook changed between February (before COVID-19 required people to work remotely) and April (after the transition to remote work); they found that weekend usage of mobile Teams increased from 25% of weekday usage to 35%, and weekend usage of PC Teams increased from 36% to 44%. Also, for SwiftKey users, weekend use of Outlook grew more than weekday usage; This may suggest that some communications are extending into weekends, and that mobile apps are seen to be more suitable for these communications.

Importantly, the observed longer workdays may be partially due to the increased interweaving of work and life tasks (see later in this chapter), meaning that the increase in actual time spent working may be smaller than the increase in the workday. However, this is unlikely to be the only factor. Prior research found that remote workers were more likely to work overtime than those with more traditional workplace arrangements [122]. Remote workers may be inclined to work more hours to signal "work devotion" in lieu of being able to do so through consistent physical presence at the office [54].

Another potential source of additional working hours is the elimination of commutes. Microsoft survey respondents reported an average total commute time of approximately one hour per day and about half of them said that not having to commute was a very important benefit of working from home [46]. In a survey of Microsoft employees in China, 59% listed commute time as a reason they continued to work from home some days after the offices opened [176].

The previous chapter on [Collaboration and Meetings](#) provided additional detail about increases in meeting time and after-hours meetings and communications.

*2.2.4 Some, but not all, employees with children struggled with childcare.*

Multiple studies reported that the additional childcare responsibilities due to COVID-19 complicated working from home [83,156]. When asked about the top factors contributing to work stress, 24% of information and firstline workers in India and 21% in Brazil selected "Balancing childcare or homeschooling with work," compared to 16% in the US, 15% in the UK, 14% in Australia, 12% in Germany and 4% in Japan [154]. In an external survey of remote workers [35], 85% of women with childcare responsibilities reported that their caregiving responsibilities were making it somewhat or much more difficult to attend to work, as did 70% of men who were caregivers [35]. The same survey also found other differences in the work-from-home experience of people with caregiving responsibilities. People with childcare responsibilities commented on exhaustion, the need to work around the clock to catch up, and challenges in homeschooling children; they were eating less healthily than before and exercising less [35]. Overall, the study found that older males without childcare responsibilities had a more positive work from home experience than those who had to balance work with caregiving responsibilities.

However, at Microsoft, the effect appeared to be smaller; fewer than 60% of employees with children reported difficulty in handling childcare responsibilities. Difficulty with childcare was correlated with self-reported productivity, but just having children was not: 43% of those who reported childcare difficulties reported their productivity had decreased while working from home compared to 29% for those who did not have children in school or daycare prior to COVID-19 and 23% for those with children who reported no difficulty with childcare. There were also cultural nuances: in India, some employees reported that staying with larger family meant having to disproportionately take on more of the household responsibilities while losing the benefits of having household help [83]. Employees with family duties also tended to report different specific challenges. Employees who previously had children in school or childcare were more likely to indicate major challenges of "More distractions or interruptions" and "Less time to complete my work", and less likely to indicate "Lack of motivation" [46]. In OXO, 5% of people consistently reported issues related to "Kids and Distractions" as the hardest part of their day for the first 7 weeks following the move to remote work [26].

*2.2.5 Managers appeared to be especially hard hit.*

An internal study [69] found that while all members of a 3,500-person org were, on average, working a bit more, managers seemed to be the hardest hit. Relative to the same period in 2019, software engineering people managers showed a 300% increase in afterhours instant messaging; 25% increase in afterhours meetings; as well as a 24% increase in completed pull requests. This suggests that people managers were both doing more to manage their teams and doing additional IC work. Another internal study found that managers represented a disproportionate amount of those employees who had experienced an increase in both workweek and collaboration hours [151].

These changes to manager collaboration appeared to be explained by a combination of remote work and the fact that they were working during a pandemic, and by comparing managers who were new to remote work with managers who were already working from home. A different study [182] found that remote work specifically increased managers' scheduled meetings by 5%. A study looking at Microsoft customers similarly found that increases in business planning and direct report reach-outs had a hefty impact on managers' collaboration hours [6]. A potentially concerning trend for managers is the possibility that this additional workload will not subside with the return to the office. In Microsoft China, where employees have begun to return to the workplace, the return-to-work period has seen manager 1:1 time climb even higher than during the fully remote period [176].

*2.2.6 The effect of remote work differed across roles and individual characteristics.*

Remote work disrupted the ways that people spent their workdays, and the changes – like the changes in reported productivity and satisfaction – differed by role [151,182]. Certain workers and types of work were more affected by the transition. According to an external survey [35], people's ability to focus was impacted by childcare responsibilities, compared to those without childcare responsibilities. There was also a gender divide – more female caregivers (26%) found it extremely difficult to focus, compared to 10% of male caregivers.

Causal inference on telemetry data [182] shows that while scheduled meetings for Microsoft US employees increased, the effect attributable to an individual working from home – as opposed to COVID-19 more broadly – is actually negative (-4%). Remote work increased unscheduled call hours (+112%), emails sent (+2.6%) and instant messages sent (+30%). For individual contributors, the total of scheduled meeting hours + unscheduled call hours remained flat, while unscheduled in-person interactions necessarily fell, implying a drop in synchronous communication time. This is consistent with prior literature [1] showing that people tend to exchange less information when moving to remote. These effects also varied across roles: they were significant for people specialized in sales, marketing, program management, and business support, but were insignificant for researchers, scientists, and software engineers [182].

The previous chapter on Collaboration and Meetings provided more information on how different types of meetings have been affected by the shift to remote work.

Individual characteristics also appeared to affect how well people were able to focus and collaborate effectively while working from home. Some challenges were particularly acute for neurodivergent professionals (such as those with autism, ADHD, learning disabilities, or psychosocial disabilities), where sensory stimuli in the home office, workspace setup, and getting into the mindset of working when surrounded by home could all make it difficult to focus [100]. Actions neurodiverse professionals took to mitigate some of the challenges included avoiding placing their desk near a window to avoid mind wandering, using timers to prevent time agnosia (inability to perceive passage of time), using fidget spinners to help remain focused, and creating spatially separate locations for work and life (e.g., not working in the bedroom) to compartmentalize their thought processes. Many neurodiverse individuals worked overtime (12-14 hours) to perform their individual focused work after combating all the attentional challenges. The flexibility that the option to work from home offers also made work more accessible to some people with disabilities, by avoiding mobility challenges around commuting, affording more flexible scheduling, and allowing people to have more ability to customize their work environments [163].

Video collaboration elicits a diverse range of accessibility issues. Because video calling (like remote collaboration technology in general) relies heavily on the visual channel, it presents obvious challenges to people who are blind or low vision. However, video calling is vital to people who rely on lip reading, especially since the pandemic response afforded capturing frontal views of each individual, which is actually better than the view one gets in a meeting in-person. It can also be challenging for neurodivergent professionals; they often prefer having their videos off as they occasionally perform certain activities to remain stimulated (e.g., fidgeting, pacing back and forth) or to calm down (e.g., petting a stuffed toy) during a meeting. These activities can be misconstrued as lack of attention or can cause distraction for other attendees. Turning off video means that they do not need to "pass" as paying attention during a meeting [100]. See Chapter 6 on the Societal Implications of remote work for additional discussion of how professionals with disabilities are being impacted.

Collaboration difficulties can also decrease the productivity of focus time. For example, in external interviews, some people reported that they had more flexibility to focus on their solo work during the day, but the focus in isolation sometimes led to misalignments and the need to redo tasks [15].

The previous chapter on Collaboration and Meetings provided more detail regarding general changes and variations in collaboration patterns.

### 2.3 Challenges and Benefits Are Closely Linked

*2.3.1 Remote work provided flexibility during the pandemic while also blurring the work-life boundary in problematic ways.*

Flexibility was reported as an important benefit of working from home across surveys in different geographies, both inside and outside of Microsoft [26]. Many people whose flexibility increased used it to interweave life and work (e.g., do a load of laundry between meetings, take time to cook a better lunch, or move more during the day) [26,46,69,128,156]. In one study, a Microsoft employee expressed appreciation for having the ability to create their own work schedule: they liked, for example, being able to split the work day into a morning block and a late afternoon block with two hours in between to recharge and take care of home duties [66].

However, as seen in prior research on remote work [1], the downside of the flexibility of remote work, particularly when it involves working from home, is the blurring of the boundary between work and home life that comes from the elimination of physical boundaries separating the office and home – as well as the temporal boundary afforded by a commute [26]; in a survey of US Microsoft employees, 72% of those who said flexibility is an important benefit also said that "lack of boundary between work and personal life" was a challenge in working from home. Information workers internal and external to Microsoft reported feeling 'always on' and having a difficult time switching off [26,128,156]. Many find it challenging to manage household chores, the immediate needs of dependents, and work meetings all at the same time in the same space [22,26,69,128,156]. A reminder that everyone is experiencing the pandemic differently from a survey commissioned by Microsoft in August 2020: a higher proportion of information and firstline workers in India, Brazil and Japan (>37%) say the lack of separation between work duties and personal obligations has negatively impacted their well-being compared to those in the US, UK and Australia (<26%) [154].

People also reported missing the time for quiet thinking and listening to podcasts that was afforded by their commutes [26]. A study with external business decision makers found that leaders and managers were aware of these challenges [62].

Unfortunately, some people taking advantage of the flexibility (e.g., to manage caregiving responsibilities during the day and work in the evening) can make it hard for others to maintain work-life boundaries: respondents in multiple surveys reported that the work patterns of collaborators can impact their ability to maintain temporal boundaries between work and personal life as increasing number of work-related emails and messages are sent after hours [26,100]. Some people reported adapting their practices to remain productive – they replaced old routines and notions of boundaries with new structures and rituals to condition for productivity, e.g., dressed up formally even while working from home to set the tone for productivity, or included micro mental breaks to help with context switching when meetings are stacked [15]. Some neurodivergent survey respondents reported trying to keep non-work related artifacts out of sight during work time to maintain the boundaries of work and non-work [100]. There was evidence of adaptation as people settle into long-term remote work in an OXO study: the number one reported challenge after 20 weeks of remote work was "work" while challenges such as "distractions" and "work-life balance" became less common [26].

See Chapter 5 on Devices and Physical Ecosystems for additional discussion of physical boundaries.

*2.3.2 Non-work distractions increased, whereas work-related distractions decreased for many.*

About half of Microsoft software engineers and program managers surveyed said they had experienced an increase in non-work related distractions (e.g., children, laundry, TV) [46]. In an external survey, over 65% of the respondents reported that they find it difficult to concentrate during remote work due to external interruptions [35]. In a different external study, people specifically reported challenges in carving out time to focus [156]. This is consistent with prior work: "Interviews with telecommuters have suggested that the ability to avoid distractions is important to being effective as a remote worker." [58]. Non-work related distractions can be particularly problematic for neurodivergent individuals in a home-office setting [100]. They strive to keep workspace and personal space completely separate so that they do not get distracted by non-work related items.

The picture for work distractions is more mixed. Studies have found an increase in IMs among Microsoft employees and information workers in an external survey reported that the messaging was disruptive [156,182]. However, 40% of US Microsoft employees surveyed said they have fewer work-related distractions while working remotely, and only 21% said they have more [46]. Microsoft employees in China also reported the ability to focus as a big benefit of remote work. In an OMEX study [181] some Microsoft employees reported that working from home allowed them to focus more, but a hybrid approach would eventually be most beneficial as some tasks might be better performed at home and other tasks are more suitable in the office (e.g., collaboration).

For neurodivergent professionals, work-related distractions can originate from the remote meeting platforms in ways that do not affect neurotypical people [100]. For example, visual background or noise from meeting partners' can be particularly distracting to neurodivergent people. Some virtual backgrounds can be distracting for neurodivergent individuals as they can get fixated on the background and lose focus on the meeting content [100]. The increasing number of notifications also create additional challenges for neurodivergent individuals; they often turn off notifications to maintain their focus on their ongoing activities [100], but that can result in missing time-sensitive information, as also shown in past work [78].

The previous chapter on Collaboration and Meetings discussed the potentially distracting role of meeting chats and other forms of multitasking. We discuss the childcare challenges specific to COVID-19 work from home later in this chapter.

*2.3.3 Many workers felt collaboratively and socially isolated.*

People are missing casual water cooler conversations and reporting increased feelings of isolation and lack of connection [5,22,26,69,128,156]. In one employee survey "Isolated", "Disconnected" and "Lonely" were amongst the top words Microsoft employees used to describe how they were feeling [128]. New hires may face heightened risk of isolation: data on software engineers shows recent new hires have fewer collaboration hours and smaller networks compared to new hires who joined Microsoft before the pandemic [69].

A study on Microsoft workers from India found that it was primarily those who were new to the workforce who thrive on the community experience and are missing working in the office the most [83]. In other internal surveys, 46% of respondents said that their needs for spontaneous interaction were not being met and a majority of respondents considered all types of social meetings to be least effective [138] . Many neurodivergent professionals do not enjoy online social meetings, because often a few people do most of the talking (often those who start the meeting) and their topic of interest may not align with everyone [100]. Also, the absence of organic side conversations in remote calls (breakout rooms are more formal) makes participation more difficult, if the main topic of conversation is not interesting [100].

In a global study by Qualtrics and SAP, 75% of people say they feel more socially isolated since the outbreak of the pandemic. In an HBR report, 1 in 3 employees say their team does not maintain informal contact (e.g., asking how each person, team, etc. is doing) while working from home. People who are lacking in informal contact are 19% more likely to report a decline in mental well-being [143]. Another survey found that people without caregiving responsibilities are feeling more isolated than those with such responsibilities [35]. Information workers often found communication tools to be a poor substitute for in-person connections [156]. Earlier in the post-COVID period many meetings included a check-in time to socialize or talk about mental, emotional and physical well-being. These have since largely gone and are being replaced by intentional meetings for this purpose, but due to video conferencing being draining, people are choosing not to attend these social meetings and losing out on the social opportunities entirely [138]. An internal study found many people use gaming to try and build social capital and are gaming more often. They also found many people (50%) are reporting doing more social capital building activities, with the most common being listening to coworkers who need to talk [138].

A study that analyzes telemetry data [182] also shows that people are more collaboratively isolated when working from home: more focused time and less meeting time. In a global Microsoft survey in April, 42% of employees reported feeling less connected (22% more) and 32% feel they have fewer opportunities to collaborate during COVID-19 (24% more); particularly employees with less prior remote work experience, those in management and those in Puget Sound and Engineering feel less connected (>50%) [128]. In a separate study, the share of individual contributors who reported favorably on their team connectiveness dropped from 90% in April to 74% in December; for managers it drop from 95% to 83% [110]. Prior work suggests that these feelings of increased isolation are not due solely to COVID-19 and quarantine. A number of studies on telecommuters prior to COVID-19 reported that remote workers feel socially isolated from their colleagues [116,135], with in-person interaction in the office being most important for maintaining workplace friendships [150]. Workers on crowdsourcing and freelancing platforms like Mechanical Turk or Upwork, where remote work is the norm, often form off-platform social networks to provide each other with companionship and social support. More generally, the evidence suggests that long-term ubiquitous remote work negatively affects relationships among coworkers and teams [1,12]. Telemetry from Surface devices around the world has shown an increase in audio website and app usage, driven largely by communication apps, while entertainment usage has decreased since the pandemic [64]. In Western Europe, we saw an increase in IMs that did not decrease when they returned to a hybrid office [151]. Across Microsoft we saw an increase in emails marked as "focus" during the beginning of the pandemic, but this has since reached a new low baseline. On the contrary, Teams usage has gone up and stayed up [166]. All of this suggests people are looking for more ways to communicate and are using a variety of methods (emails, Teams, Zoom, WhatsApp, etc.) to do so while remote.

*2.3.4 Remote work provided some benefits for worker health, but also many challenges, particularly with regard to well-being.*

In addition to increased flexibility, working from home has created other quality-of-life benefits for many. Less time spent commuting, more time spent with family, reduced health risks from the pandemic, and improved diets were some of the benefits mentioned in internal surveys [26,46,128]. Some people also reported more opportunities for micro-exercise (e.g., yoga during Teams meetings with the camera off or walking 1:1s over the phone), but others felt they had become more sedentary [26]. In the OMEX study [181] some people reported that inclusivity had improved as team members were more tolerant of differences in working style.

In contrast with the above benefits, some employees and external information workers reported feelings of stress and a feeling of always working [26,128,156]. About 85% of the respondents in an external survey of US-based respondents including technology professionals, academics, marketing and public relations professionals said that they feel nervous and stressed [35]. People are also reporting being more overworked as remote work continues [26]; this may be aggravated by people not taking vacation since COVID-19 means they cannot travel and reporting having a harder time getting back into work when they do [26]. Survey respondents and researchers question whether the current work pace is sustainable and raise concerns about long-run burnout [26]. A couple of days in May were designated "COSINE Health Days"; it seems that about two-thirds people actually took the days off and many were appreciative: "I didn't realize how badly I needed a break" [69].

A global study of over 2,700 employees across more than 10 industries undertaken by Qualtrics and SAP during March and April 2020 showed that since the outbreak of the pandemic, 65% of people report higher stress, 57% are feeling greater anxiety, and 53% say they feel more emotionally exhausted [143]. The same study showed that 40% of people at every seniority level of a company have seen a decrease in mental well-being [143], while another study of 1200 US full-time employees conducted by Ginger, the leader in on-demand mental healthcare, showed that 70% of workers said the pandemic has been the most stressful time in their careers [81]. From a survey commissioned by Microsoft in August 2020, over 30% of information and firstline workers say the pandemic has increased their sense of burnout at work. This number jumps to 44% in Brazil and drops to 10% in Germany compared to 31% in the US. The same survey found that causes of workplace stress differ between remote and firstline workers: lack of separation between work and life and feeling disconnected from co-workers are the top 1 and 2 stressors for remote workers, while the inability to socially distance/worry about getting COVID-19 and unmanageable workload/workhours are the top 1 and 2 stressors for Firstline workers. 70% of information and firstline workers surveyed said they think meditating could help them decrease their work-related stress. Again, this number jumps to 92% in India and 84% in Brazil, compared to 73% in the US and 53% in the UK [154].

Challenges in having to adapt in-person practices through new digital tools that may result in broken workflows also contribute to mental stress and exhaustion. Survey respondents reported that technical challenges resulted in more annoyance and frustration compared to in-person work, as information workers were relying so heavily on digital tools to get their work done [156].

In addition to mental stress, there is the additional physical stress on the body from remaining in one place, as people no longer need to walk between offices, or walk as far to the restroom or to get coffee, etc. [26,46,128,149,163], and this is causing physical pain [26,128]. People are complaining of their eyes being sore from staring at the screen; stiffness; aches; carpel tunnel pain; and back issues [26]. In a study of external workers, 90% reported spending more time on their electronic devices [35]. People are reporting not moving for an entire day except for bio breaks or to get food, and this is concerning knowing the negative effects of spending most of the day sitting (increased risk of high blood pressure, blood sugar concerns, cancer, and even premature mortality) [1]. The decrease in movement may also be aggravated by COVID-19 causing gym closures – in one survey 75% of respondents said they were getting less exercise than prior to the pandemic [35]. The potential long-term health implications of ubiquitous remote work are serious but under-researched [1].

Lastly, brain scan-based studies provide some evidence that virtual meetings require more cognitive load, but that the Teams Together mode may reduce load relative to the traditional grid view and could help eliminate online meeting fatigue [79]. Another study found that an "attentional spotlight" of the audience based on their affective responses provided presenters with a better notion of how their talk was going, and where there was agreement or questions/confusion relative to the current Teams' grid view of the audience or a randomly chosen spotlight of an audience member [118]. This reduced the cognitive burden and stress from not being able to "read the room" or know how your presentation was accepted. The previous chapter on Collaboration and Meetings discussed how the increase in virtual meetings added to both physical and mental stress.

Chapter 5 on Devices and Physical Ecosystems describes more research on the specific physical effects of working from home.

*2.3.5 Employees worried that their hard work would not be visible to their managers.*

Microsoft employees and external information workers have expressed concerns about the need to "look productive" and the fact that their work is now less visible to their manager or team [26]. This may be aggravated by job security concerns resulting from COVID-19's effect on the economy, but it is also expected from the prior literature on remote work: physical presence in the office is a typical means of signaling "work devotion," and remote workers before the pandemic frequently felt the need signal devotion in other ways (e.g., by working longer hours as discussed earlier) [54,122]. In this vein, while some Microsoft employees appreciate the benefit, some have indicated reluctance to take the Paid Pandemic School and Childcare Closure Leave [26,69]. One benefit of a stronger desire to see and be seen may be that people are being more open to ask for help and there are more open discussions [15].

While employees are being more demonstrative about work due to concerns about job security, a survey of external business decision-makers found that leaders want to use KPIs and project completion to evaluate employees and are not interested in monitoring employee behavior [62], suggesting that at least in some cases, employee concerns about looking productive may be unwarranted. These leaders are already speculating about how to best assess productivity and performance in a new "flexi-work" or "hybrid" work environment [62].

*2.3.6 Conversely, there was increased concern about how their work did show up.*

While people worried their work would not be visible, they also expressed increased concern about the fact that traces of their work often persisted. Because technology mediates more of the work a person does when working remotely, more interactions can end up being recorded, including written text (e.g., email or IM) and video (e.g., remote meetings). This made some employees more hesitant to ask potentially stupid questions while working remotely, and when they did ask questions they sometimes spent a lot of time on wordsmithing to make sure the question was clear (since clarifying can harder when not co-located) and made them look good [26].

> *"Working from home also takes away the easy access to colleagues nearby for a quick question or a "hallway meeting". Yes, the tools to attempt to replace that are available but aren't perfect replacements. Everything is recorded in writing (chats) or I'd have to pick up the phone and call which takes a bit more effort so I'm just less likely to do it." [160]*

The notion of "remote work as surveilled work" gained prominence during the ubiquitous work-from-home period [145]. Prior literature on remote work has documented employers' desire to ensure workers are working when expected [1], but also identified negative effects of electronic performance monitoring, including on employee productivity, creativity, trust and well-being [11].

**2.4 Looking Forward**

*2.4.1 Roles of managers and leaders are evolving to adapt the challenges of remote work.*

There is preliminary evidence that managers may be able to help with any workday length and productivity challenges: Microsoft employees who received prioritization support from their managers were 2.5 times more likely to report maintaining their productivity levels and work-life balance in comparison to those who did not receive prioritization support [110]. A study of telemetry of one organization in the company found that employees who had the most 1:1 time with their managers saw a much smaller increases in weekly working time (1.5 hours vs. 3 hours) [4]. Similarly, a study of Microsoft employees in Denmark found that the average meeting and email load decreased as managers' 1:1 time increased. The inverse relationship suggests 1:1 time enabled managers to help employees prioritize work and resolve issues faster [151]. Research has also shown employees who have more one-on-one time with their managers are less likely to be disengaged [121]. Leadership in OXO instituted a policy to increase manager check-ins, which had a positive effect on 41% of survey respondents (4% negative) [26].

Managers are critical points of connection in terms of providing holistic support to employees – reducing negative impact on physical and mental well-being, maintaining morale over reduction in force and so on, suggesting soft skills of managers are becoming even more important than ever before [62]. Recent research of external decision-makers has shown that the roles of managers and leaders are also evolving with the changes that remote work brings: leaders are moving faster to make decisions to take quick action on maintaining business continuity and stability, but with the potential negative consequences of bypassing regular checks that might emerge downstream [62].

Alignment of the lived experience of leaders and managers is important to create a working environment that caters to the needs of employees with different challenges. Senior leaders recognize their role in creating a more democratized and humanized workplace and set examples of how they show up on video from home as well as appearing to be more accessible online compared to in person [62].

A meta-analysis of past research showed a positive correlation between remote work outcomes and manager relationship quality, but since it was cross-sectional the association could be due entirely to the fact that (under normal circumstances) managers may only allow remote work for employees with whom they have a good relationship [51].

*2.4.2 Where to work once it is safe? Workers report mixed preferences but lean toward a hybrid model.*

Multiple surveys have found that, for the post-COVID workplace, most employees want a hybrid model of work in which people work part-time in the office and part-time from home.

In Microsoft China, where offices have reopened, a large majority of respondents to a survey preferred a hybrid model of work (69%) compared to purely remote work (19%) or the traditional office model (11%) because it enabled people to combine the advantages of both remote and in office [176]; the same survey found that 69% of people thought that they were most productive in a hybrid work model. These numbers align closely with those of a similar survey from COSINE, in which 65% of employees said they wanted a hybrid model post-COVID (vs. 26% remote-only and 9% traditional) [69]. Additionally, from the CSEO Global Employee Satisfaction Survey (GESS) in November 2020, 63% of Microsoft employees ideally want a hybrid model compared to 35% fully remote and 3% never remote [128]. Similarly, 66% of Google employees want a hybrid model compared to 20% fully remote and 10% never remote [94] For some, these preferences are in spite of concerns that a hybrid model will be harder to manage and potentially create inequality between people in the workplace and people working remotely [26].

There are differences in how people think hybrid work will work best. In a poll regarding same-room vs. hybrid vs. remote meetings, Microsoft study participants with prior remote work experience were much less likely to support "Prioritize same room meetings for which planning is difficult," suggesting maybe remote workers underestimate the value of co-location in resolving ill-defined tasks and office workers overestimate it [138]. On the other hand, there was near-universal agreement on "Prioritize hybrid meetings in which all attendees engage fully" and "Prioritize same-room meetings which are well-planned so that meetings are efficient."

Results from outside Microsoft also reveal a preference for hybrid work, although the traditional model does better in these findings. A survey of US workers found that, post-COVID, 24% would like to work from home all of the time, 56% would like to work from home at least some of the time, and 20% would like to return to the office full-time [18]. Interestingly, the preferences of company leaders with regards to hybrid work likely differ from those of their employees. A survey of firms found that, on average, they only want the percent of working days worked from home to increase from 5.5% pre-COVID to 16.6% post-COVID [2].

All of the above findings refer to workers' self-reported preferences for a biologically-safe post-COVID world. In the short-term, people are less positive about returning to the office, whether that is in a hybrid or in a traditional model. In May 2020, Microsoft employees expressed concerns, anxiety, and even fear around returning to the office given the challenges of social distancing. Some specifically mentioned they will not return to the workplace or travel until there is a COVID-19 vaccine. In November 2020, despite soft openings across many regions, 94% of Microsoft employees continued to work from home full-time (80% across Asia region) [128]. In an external survey of information workers, 78% of respondents reported that they did not want to return to the workplace yet and that cleanliness of the workplace is an important factor for them in deciding whether they want to return [35].

Even pre-COVID, the desire for location flexibility was prominent among young people. In a global external survey of youths aged 16-25, 72% agreed that "My ability to choose how and when I work impacts how satisfied I am with my job" (11% disagreed). In the same survey, 63% agreed that "I can be a productive and effective employee for an organization without going to the workplace every day" (18% disagreed) [103]. However, many young people rely on the office for socializing as well, meaning hybrid may still be preferred by them to fully remote work [56].

*2.4.3 Business decision-makers are considering long-term changes.*

Chapter 6, which covers Societal Implications, will discuss some long-term implications of companies planning for a future with more remote work, including changes to real estate footprint and use of freelancers. Some business leaders are being empathic in deciding who should return to work with special consideration of the physical and mental well-being of the employees [62]. However, planning for the hybrid model is logistically complex – and reorganizing the physical environment to support distancing needs was identified as a goal by business leaders [62] and as a need by employees [35]. Employee well-being is quickly becoming a key priority for employers – listed as one of the top 5 business priorities in FY21Q1 by roughly a third of Microsoft's global customer and partners [33]. There is a latent desire among IT and Business decision makers for metrics to provide a broad understanding of the well-being, productivity and collaboration of the employee base – without intruding on privacy [42].

71. Shawn Hubler, Thomas Fuller, Anjali Singhvi, and Juliette Love. 2020. Many Latinos Couldn't Stay Home. Now Virus Cases Are Soaring in Their Communities. *The New York Times*. Retrieved from https://www.nytimes.com/2020/06/26/us/corona-virus-latinos.html
72. Ginger Hudson. 2020. *ITDM in Commercial: Post-pandemic device considerations*. Microsoft (Internal).
73. Ginger Hudson. 2020. *Device Hygiene Study*. Microsoft (Internal).
74. Ginger Hudson. 2020. *Workstation Evolution During the Pandemic*. Microsoft (Internal).
75. Ginger Hudson and Steven Derhammer. 2020. *Devices and Remote Work Study*. Microsoft (Internal).
76. Lori Ioannou. 2020. 1 in 4 Americans will be working remotely in 2021, Upwork survey reveals. *CNBC*. Retrieved January 21, 2021 from https://www.cnbc.com/2020/12/15/one-in-four-americans-will-be-working-remotely-in-2021-survey.html
77. Shamsi Iqbal and Kirk Daues. 2020. *Enterprise Customer Adaptation to Remote Work and Planning for a Hybrid Future*. Microsoft (Internal).
78. Shamsi T. Iqbal and Eric Horvitz. 2010. Notifications and awareness: a field study of alert usage and preferences. In *Proceedings of the 2010 ACM conference on Computer supported cooperative work - CSCW '10*, 27. https://doi.org/10.1145/1718918.1718926
79. Jared Spataro. Reimagining virtual collaboration for the future of work and learning. Retrieved from https://www.microsoft.com/en-us/microsoft-365/blog/2020/07/08/reimagining-virtual-collaboration-future-work-learning/
80. Karen A. Jehn and Priti Pradhan Shah. 1997. Interpersonal relationships and task performance: An examination of mediation processes in friendship and acquaintance groups. *Journal of Personality and Social Psychology* 72, 4: 775–790. https://doi.org/10.1037/0022-3514.72.4.775
81. Jen Porter, Bernie Wong, and Kelly Greenwood. 2020. How to Form a Mental Health Employee Resource Group. *Harvard Business Review*. Retrieved from https://hbr.org/2020/05/how-to-form-a-mental-health-employee-resource-group
82. Anya Kamenetz. 2020. Families Of Children With Special Needs Are Suing In Several States. Here's Why. *NPR*. Retrieved from https://www.ijpr.org/education/2020-07-23/families-of-children-with-special-needs-are-suing-in-several-states-heres-why
83. Prerrna P. Kapoor, Abhijit Bairagi, and Anita Isola. 2020. Impact of COVID-19 crisis on the future of work in India. In *Microsoft New Future of Work Symposium*. Retrieved from https://www.microsoft.com/en-us/research/publication/impact-of-covid-19-crisis-on-the-future-of-work-in-india/
84. Harmeet Kaur. 2020. Why rural Americans are having a hard time working from home. *CNN*. Retrieved from https://www.cnn.com/2020/04/29/us/rural-broadband-access-coronavirus-trnd/index.html
85. Kristi Kelly and Darren Austin. 2020. *Mental Wellbeing*. Microsoft (Internal).
86. Dawn Klinghoffer, Candice Young, and Dave Haspas. 2019. Every New Employee Needs an Onboarding "Buddy." *Harvard Business Review*. Retrieved January 21, 2021 from https://hbr.org/2019/06/every-new-employee-needs-an-onboarding-buddy
87. Dawn Klinghoffer, Candice Young, and Xue Liu. 2018. To Retain New Hires, Make Sure You Meet with Them in Their First Week. *Harvard Business Review*. Retrieved January 21, 2021 from https://hbr.org/2018/06/to-retain-new-hires-make-sure-you-meet-with-them-in-their-first-week
88. Andrew J. Ko. 2019. Why We Should Not Measure Productivity. In *Rethinking Productivity in Software Engineering*, Caitlin Sadowski and Thomas Zimmermann (eds.). Apress, Berkeley, CA, 21–26. https://doi.org/10.1007/978-1-4842-4221-6_3
89. Ella Koeze and Nathaniel Popper. 2020. The Virus Changed the Way We Internet. *New York Times*.
90. Justin Lavelle. 2020. Gartner CFO Survey Reveals 74% Intend to Shift Some Employees to Remote Work Permanently. *Gartner*. Retrieved January 17, 2021 from https://www.gartner.com/en/newsroom/press-releases/2020-04-03-gartner-cfo-surey-reveals-74-percent-of-organizations-to-shift-some-employees-to-remote-work-permanently2
91. Heather Layne and Giancarlo Cozzi. 2020. The rise of the 30-minute meeting. *Home | Work Blog from Microsoft Workplace Insights*. Retrieved from https://insights.office.com/workplace-analytics/the-rise-of-shorter-meetings-and-other-ways-collaboration-is-changing-with-remote-work/
92. Kyung-Hee Lee and Ellen Ernst Kossek. 2020. *The coronavirus & work–life inequality: Three evidence-based initiatives to update U.S. work–life employment policies*. Behavioral Science and Policy Association.
93. Lydia Lee. 2020. Post-COVID, More Office Designs Include Permanent Outdoor Workspaces. *Metropolis*. Retrieved January 21, 2021 from https://www.metropolismag.com/architecture/workplace-architecture/post-covid-outdoor-workspaces/
31